\documentclass{llncs}
\usepackage[pdftex]{graphicx}
\usepackage{flushend}
\usepackage[tight,footnotesize]{subfigure}

\begin{document}
\author{Nikos Fotiou \and  Vasilios A. Siris \and George C. Polyzos}
\title{Interacting with the Internet of Things using Smart Contracts and Blockchain Technologies}
\institute{Mobile Multimedia Laboratory, Department of Informatics\\
School of Information Sciences and Technology\\
Athens University of Economics and Business\\
76 Patision, Athens 10434, Greece\\
\email{\{fotiou, vsiris, polyzos\}@aueb.gr}
}
\maketitle

\abstract{
Despite technological advances, most smart objects in the Internet of Things (IoT) cannot be accessed using technologies designed and developed for interacting with powerful Internet servers. IoT use cases involve devices that not only have limited resources, but also they are not always connected to the Internet and are physically exposed to tampering. In this paper, we describe the design, development, and evaluation of a smart contract-based solution that allows end-users to securely interact with smart devices. Our approach enables access control, Thing authentication, and payments in a fully decentralized setting, taking at the same time into consideration the limitations and constraints imposed by both blockchain technologies and the IoT paradigm. Our prototype implementation is based on existing technologies, i.e., Ethereum smart contracts, which makes it realistic and fundamentally secure. 
}


\section{Introduction}
The Internet of Things (IoT) is an emerging paradigm that has already attracted the attention of both academia and industry. The IoT is expected to penetrate various aspects of our life, allowing the creation of cyber-physical applications that will improve our living conditions by enabling healthier and cheaper agricultural products, smarter energy production and consumption, safer transportation, better entertainment and wellness activities, and innovative services. The IoT will be composed of smart devices and protocols that will allow human-to-device and  device-to-device interactions. Nevertheless, these devices--henceforth simply referred to as Things--as well as the mainstream IoT use cases, present some limitations and particularities that create the need for new, innovative interaction protocols. In particular, Things will be far less powerful than traditional Internet clients and servers. Furthermore, Things will not always be connected to the Internet, e.g., in order to preserve energy or because they will be physically located in places where Internet access is not possible (at times). Finally, it should be easier for a malicious user to tamper with a Thing, hence Things should not be used for storing ``important'' secrets or for processing very sensitive information. To this end, in this paper we design, implement and evaluate a solution that enables secure interaction with the IoT by leveraging blockchain technologies and smart contracts.

A blockchain is an append-only ledger of transactions distributed throughout a network. Transactions are validated by a number of network nodes and are added in the ledger upon consensus, assuring this way that no single entity has control over the ledger. A smart contract is a distributed application that lives in the blockchain. Users can interact with a smart contract by sending transactions to its ``address'' in the blockchain. For any interaction  with a smart contract, all operations are executed by the blockchain, in a deterministic and reliable way. Smart contracts can verify blockchain user identities and digital signatures and they can perform a number of operations. The code of a smart contract is immutable and it cannot be modified even by its owner/creator. Moreover, all transactions sent to a contract are recorded in the blockchain.

Although, blockchains and smart contracts--henceforth simply referred to as Distributed Ledger Technologies (DLTs)--are considered a ``democratic'' way for maintaining transactions~\cite{ibm2014} and are envisioned to provide novel security mechanisms~\cite{Pol2017}, they have some properties that limit their (direct) applicability in the context of the IoT. Firstly, interacting with a DLT involves some computationally intensive security operations (e.g., the creation of a digital signature). Secondly, DLTs require users to maintain a private key: this key is an important secret that protects the assets of the users stored in the blockchain. Thirdly, information stored in smart contracts is public, hence smart contracts cannot be used for storing, e.g., user credentials, access control policies, etc. Similarly, information stored in smart contracts is immutable and all interactions with a smart contract are recorded in the blockchain, hence it is trivial for a third party to deduce, for example, all modifications to an access control policy.  Finally, smart contracts cannot directly interact with the physical world: the execution of a smart contract relies solely on information stored in the blockchain. 

In this paper, we design and build a solution that allows users to securely interact with the IoT using DLTs even if Things are not connected to the Internet continuously or directly. Our solution, which is built using the Ethereum transaction ledger~\cite{Wood2014}, takes into consideration the limitations and particularities of the IoT and the DLTs, is secure and realistic. With our approach we make the following contributions:
\begin{itemize}
\item	We enable access control, Thing authentication, and payments in a decentralized, secure, and efficient way.
\item	We build on existing technologies and do not propose a new blockchain, neither yet another specification for smart contracts.
\item	We preserve end-user privacy (to the degree that it is preserved by the specific blockchain used).
\item	We assure that Things are oblivious to the existence of the blockchain, do not store any blockchain-specific secret and the underlay blockchain technology is completely transparent to the Things.
\end{itemize}

\section{System Overview}
Our solution leverages our previous work, published in~\cite{Fot2016b}, that allows a Thing and an authorized user to establish a shared, session specific secret key; this key can be used for securing (using symmetric encryption) all message exchanges. This operation is achieved with the help of a third party, referred to as the \emph{Access Control Provider}. From a very high perspective, the solution described in~\cite{Fot2016b} operates as follows. ACPs maintain a user management system, as well as access control policies, associated with a (Thing provided) resource. Furthermore, each Thing shares a unique key with each ACP that handles access to its resources. Whenever a user requests a protected resource, the Thing generates a \emph{token} and sends it back to the user. The token is sent in plaintext over an unsecured communication channel: mechanisms (not detailed in this paper) make sure that any message modification, replay and man in the middle attack can be detected. ACPs and Things can calculate a new secret key, referred to as the \emph{session key} using a secure keyed-hash message authentication code (HMAC) with inputs the shared secret key and the generated token. ACPs are responsible for authenticating users and for securely transmitting the session keys to the authorized ones.

The solution described in~\cite{Fot2016b} assures that the session keys calculated by an ACP and a Thing are the same if (a) the user is interacting with the real Thing, (b) the user is authorized to access the resource, (c)~the user has not lied about his identity, and (d) no messages have been modified. Otherwise, the calculated session keys will be different, hence it will not be possible for the user to communicate with the Thing. In other words, this solution offers Thing and user authentication, user authorization, message integrity protection, and session key agreement. Furthermore, this solution has two notable properties: (a) the Thing does not have to be able to communicate with the ACP (as a matter of fact the Thing can be completely disconnected from the rest of the world) and (b) the ACP does not have to be aware of the services provided by the Things, i.e., an ACP and the \emph{service provider} can be two distinct entities.

In this work we consider a similar setup with the addition that users have to make some form of payment (not necessarily monetary) to the service providers--henceforth they will be simply referred to as providers--every time they interact with a protected resource.  In order to give a better overview of our system we present the use case of a ``smart coffee machine.'' In this use case, a smart coffee machine is installed in a shared kitchen of a building where the offices of many companies are located. Users interact with the coffee machine using their mobile phones and Wi-Fi direct. The coffee machine operator has come to an agreement with one of the companies located in that building, Company A, and each employee of that company is offered 300 free cups of coffee per year. Every time an employee of Company A wishes to order a coffee the following process is followed. The employee sends a request to the coffee machine, the coffee machine sends a token, the employee authenticates with the ACP of Company A\footnote{For simple access control policies, e.g., lists of blockchain specific public keys, this authentication process can take place over the blockchain, otherwise, further information has to be exchanged using an off-chain communication channel.} and receives the session key, the employee pays the coffee machine operator (the first 300 times 0.00 EUR and then with the value of the coffee), the employee sends a coffee request encrypted with the session key, and finally the coffee machine sends a receipt encrypted with the session key and disposes the coffee. All interactions among the user, the ACP, and the coffee operator (but not between the user and the coffee machine) utilize a smart contract, stored in a blockchain. Our system achieves the following:

\begin{itemize}
\item \textbf{Low complexity}.  Coffee machines are oblivious about the existence of the blockchain and perform only some very lightweight operations. ACPs are not aware of the services the coffee machine operator offers, neither do they have to handle payments. The coffee machine operator does not have to be aware of the user management system of Company A.
\item \textbf{Support for payments}. A smart contract makes sure that users have the necessary amount of money required for an order. Furthermore, the same contract makes sure that all payments are made prior to placing the order.
\item \textbf{User privacy protection}. No user personal information is stored in the blockchain. Similarly, coffee machines learn nothing about users. 
\item \textbf{Endpoint authentication}. A smart contract makes sure that a user is authenticated and that the ACP and the coffee machine indeed share a secret key \emph{before the user places the order}, by utilizing the session key.
\end{itemize}
Our system does not provide any guarantees for the interactions that take place in the physical world, e.g., in our use case, our system does not guarantee that the coffee machine does deliver the requested coffee.
However, the interaction and payment is recorded in the blockchain (in an immutable way), which can be used as proof in court, if it comes to that.
     
\section{System Design}
\subsection{Preliminaries and notation}
In our system, service providers and users own a blockchain specific public/private key pair. We refer to the public key of a user as $P_{user}$, and to the encryption of a message $m$ using
% GCP
(the private key corresponding to)
$P_{user}$ as $E_{user}(m)$. For simplicity, we assume that an ACP knows all $P_{user}$ of its users and all access control policies are based on these keys. ACPs, access control policies, smart contracts, and resources are identified by a URI. We refer to a URI of an entity as $URI_{entity}$. Smart contracts implement functions, which can be invoked using transactions, and generate events; $f(x,y,z)$ denotes the invocation of a function $f$ with arguments $x,y,z$, and $E(x,y,z)$ denotes an event $E$ with arguments $x,y,z$. Our system uses a keyed-hash message authentication code (HMAC), as well as a simple hash function. We refer to the digest of a message $m$ using an HMAC function $H$ and a key $k$ as $H_k(m)$ and to the hash of a message $m$ as $H(m)$. As already discussed, an ACP and a Thing end up generating a session key. We refer to this key as $sk$ and to the encryption of a message $m$, using $sk$ and a symmetric encryption algorithm as $C_{sk}(m)$. For each user $P_{user}$ there is a cost for accessing a resource $URI_{resource}$. This cost is known to the smart contract. Similarly to users, each ACP owns a blockchain specific public/private key pair denoted by $P_{ACP}$.   

\subsection{Protocols}
\subsubsection{Set up}
The protocols described in the following assume a setup phase. During this phase, the smart contract is configured with the available $URI_{resource}$ and the corresponding $URI_{policy}$ and $P_{ACP}$. For simplicity
of presentation
it is assumed that each $URI_{resource}$ is protected by a single $URI_{policy}$ provided by a single $P_{ACP}$.
\subsubsection{Straw man approach}
\begin{figure}
\begin{center}
\includegraphics[width=0.90\linewidth]{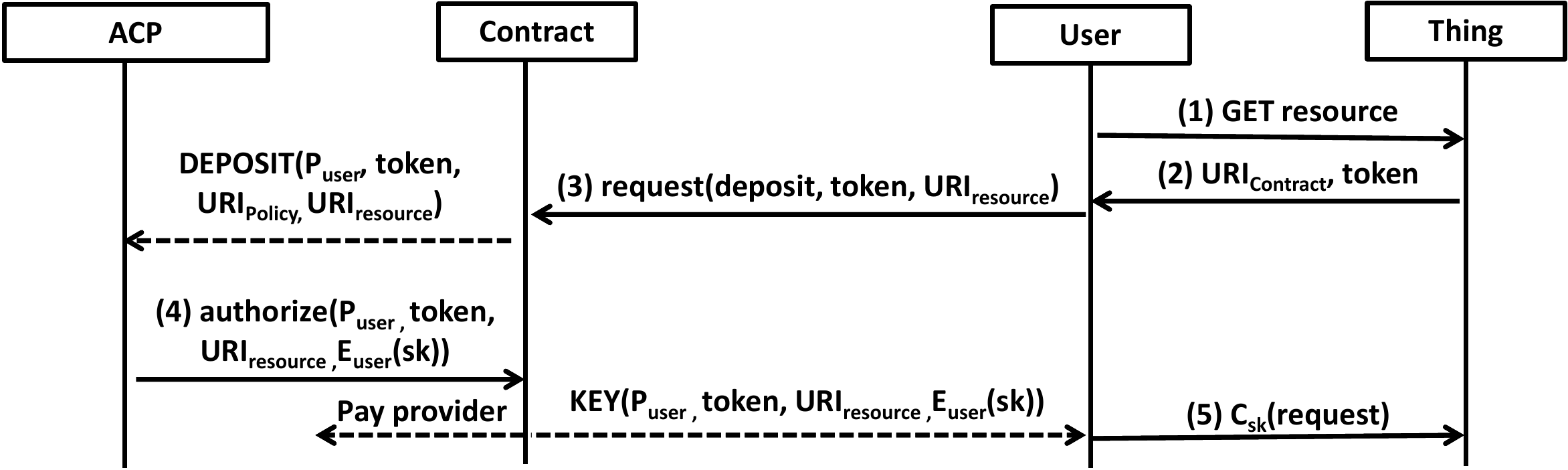}
\caption {The straw man protocol.}
\label{fig:1}
\end{center}
\end{figure}

Firstly, we present a simple protocol that implements our solution. This protocol is illustrated in Figure~\ref{fig:1}. This protocol is based on a smart contract that provides the following methods:
\begin{itemize}
\item \textbf{request}(deposit, token, $URI_{resource}$): Examines if the deposit of the user suffices for accessing the resource $URI_{resource}$. If this is true, it creates a DEPOSIT event with arguments, $P_{user}$, token, $URI_{policy}$, and $URI_{resource}$.
\item \textbf{authorize}($P_{user}$, token, $URI_{resource}$,$E_{user}(sk)$): Transfers the deposit that the user $P_{user}$ made (when she invoked the request method) to the service provider.  Then it creates a KEY event using the method input parameters as arguments.
\end{itemize}
With this protocol, initially, a user $P_{user}$ requests a protected resource from a Thing and the Thing responds with a token (generated using the process described in~\cite{Fot2016b}) and the URI of a smart contract that protects the requested resource. Then, the user invokes the \emph{request} method of the smart contract. The DEPOSIT event is broadcast and received by the appropriate ACP which  examines if $P_{user}$ can be authorized to access $URI_{resource}$. If this is true, the ACP generates the session key $sk$ (using the process described in~\cite{Fot2016b}), encrypts it using $P_{user}$,  and invokes the \emph{authorize} method of the smart contract.  The smart contract examines if the ACP that invoked the \emph{authorize} method is allowed to do so. This check is simply implemented by examining if the public key of the entity that invoked that method is equal to the $P_{ACP}$ of the legitimate ACP.  

The drawback of the straw man approach is that the payment to the provider takes place without any check.
Note that with the solution described in~\cite{Fot2016b}, the user is able to perform certain verifications \emph{after} trying to use the received $sk$. However, with the straw man approach, these verifications can only be used for a dispute resolution. 

\subsubsection{A first construction}
\begin{figure}
\begin{center}
\includegraphics[width=0.90\linewidth]{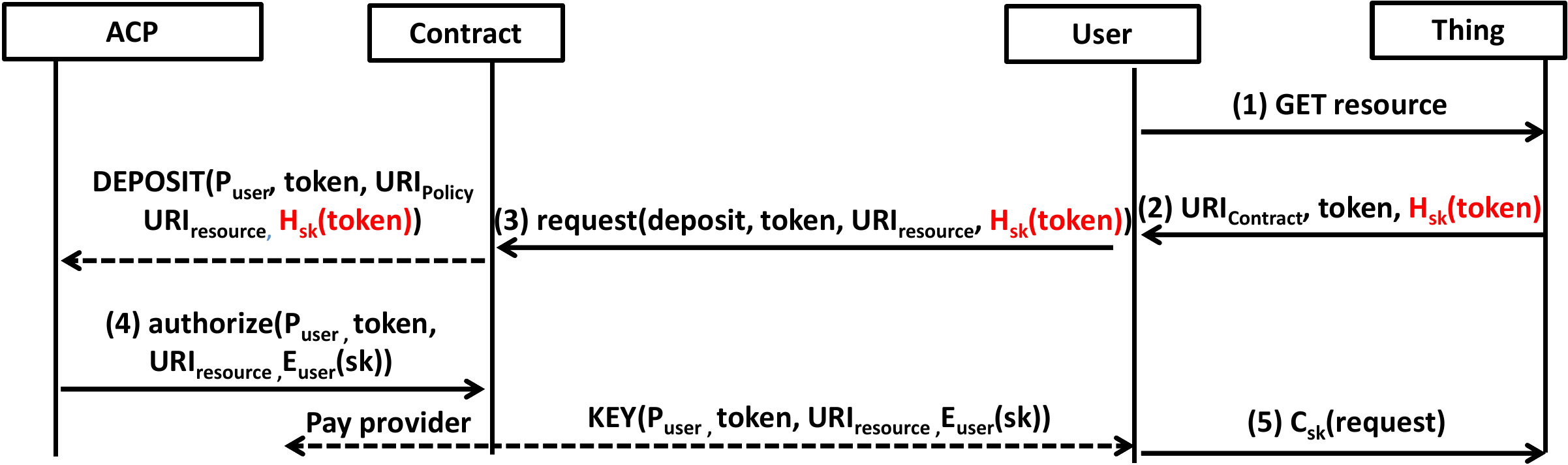}
\caption {Our first construction.}
\label{fig:2}
\end{center}
\end{figure}

We now present an improvement to the straw man approach by allowing an ACP to verify that a user is communicating with a legitimate Thing (Figure~\ref{fig:2}). In order to achieve this goal, we extend the \emph{request} method of the smart contract to include an additional field, i.e., $H_{sk}(token)$. The value for this field is provided by the Thing, in its response to a user request. Furthermore, we extend the DEPOSIT event to include this field. Now an ACP, after generating the $sk$, calculates $H_{sk}(token)$, and checks if the value of the latter calculation is equal to the value provided by the Thing. If this is true, then the Thing is considered legitimate.  

\subsubsection{A second construction}
We now extend our previous construction  to enable smart contracts to verify the relationship between a Thing and an ACP, i.e., the contract can verify that the Thing and the ACP indeed share a secret key. This functionality is achieved by having the user ``challenging'' the Thing during her request. The \emph{challenge} used is a random number, which the Thing should obfuscate in a way that only an ACP that shares a secret key with the Thing could read. The smart contract should therefore learn the challenge from the user and should expect it from the ACP. In order to ``hide'' the challenge we leverage a hash function using the process described below. 

The Thing responds to a challenge with $H(H_{sk}(challenge))$. Given a challenge, only an entity that can generate the session key $sk$ can calculate $H_{sk}(challenge)$.
Note that, in addition to the Thing, this key can be calculated by the ACP that protects the resources stored in that Thing.
Furthermore, given $H_{sk}(challenge)$ any entity, including the smart contract, can easily calculate $H(H_{sk}(challenge))$ (but the reverse process is not possible due to the properties of the hash functions). Hence, the \emph{request} method is extended to include $H(H_{sk}(challenge))$ and the \emph{authorize} method is extended to include $H_{sk}(challenge)$. Then, the smart contract can calculate the hash $H_{sk}(challenge)$, received by the ACP, and compare the output to the hash value it received from the user. If both hash outputs are the same, the contract sends the $KEY$ event. 

\begin{figure}
\begin{center}
\includegraphics[width=0.90\linewidth]{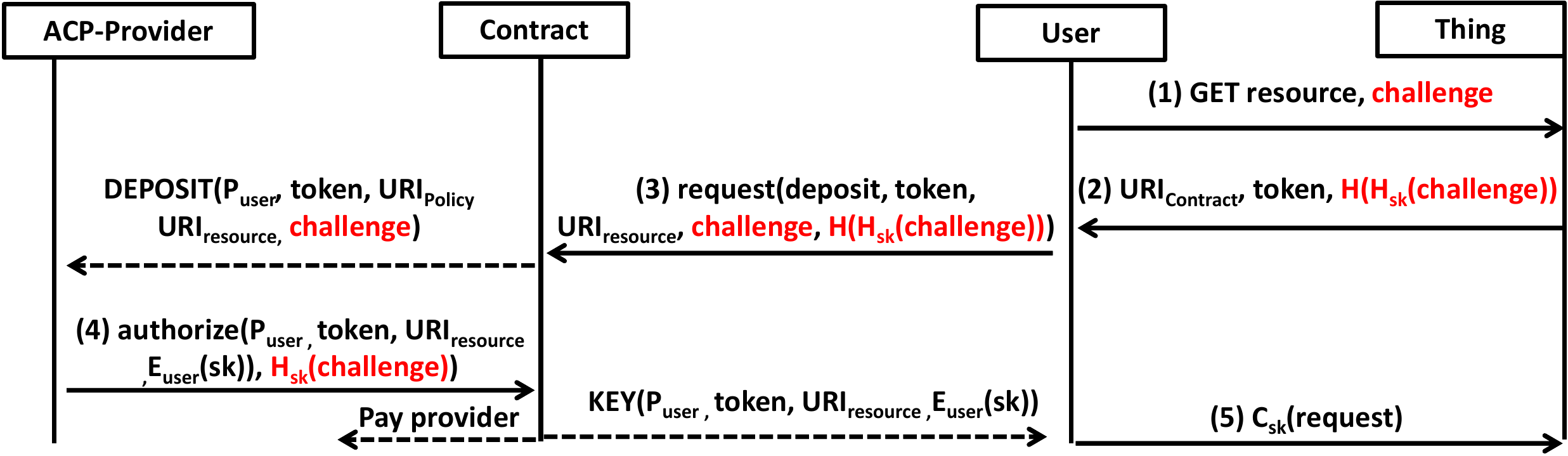}
\caption {Our second construction.}
\label{fig:3}
\end{center}
\end{figure} 

\section{Implementation and evaluation}
We have implemented the presented solution using Ethereum smart contracts.\footnote{Source code of our implementation can be found at:~https://github.com/SOFIE-project/spiot} 
This technology has some limitations that have led us to certain design choices. In particular, although each user in Ethereum owns a public/private key pair, a smart contract has access only to each user's ``address'', i.e., the last 20 bytes of the hash of her public key. This means that users have to explicitly include their public keys with every smart contract function invocation; in our implementation we have added an additional field in each function which is used for storing callee's public key. Furthermore, Ethereum keys are constructed using the secp256k1 elliptic curve; encrypting content using this curve can be cumbersome since specialized constructions, such as the  elliptic curve integrated encryption scheme~\cite{Sho2001}, are required. For these reasons we have selected to not use Ethereum's keys in our constructions, but instead we are using keys based on the Curve25519 elliptic curve~\cite{Ber2006}. Curve25519 is a well supported, fast curve which is ideal for key establishment, as it allows a user $A$ to generate a symmetric encryption key that can be used for communicating with a user $B$, using only $B's$ public key.

The main constructions of our smart contract, which is deployed in a local testbed, are implemented in five functions: $requestS()$,$request1()$, $request2()$, each implementing the $request()$ method for our three protocols (straw man, first construction, and second construction), and $authorize1()$ and $authorize2()$, that implement the $authorize()$ method for the first two protocols and for the last protocol respectively. The table below illustrates the cost, measured in Ethereum ``gas,'' for invoking each function.

\begin{table}[h]
\centering
\begin{tabular}{ | l | c |}
\hline
\textbf{Function} & \textbf{Cost measured in gas} \\ \hline
requestS()        &  123.186                             \\
request1()        &  128.218                            \\
request2()        &  253.488                             \\
authorize1()      & 57.950                              \\
authorize2()      & 63.746                              \\ \hline
\end{tabular}
\caption{Cost for invoking smart contract functions}
\end{table}

Endpoints are implemented using JavaScript. Interactions with the Ethereum blockchain are implemented using the Ethereum JavaScript API,\footnote{https://github.com/ethereum/web3.js/} whereas cryptographic operations are implemented using the TweetNaCl library.\footnote{https://tweetnacl.js.or}    
     
\section{Related work}
Prior work on blockchain-assisted access control has proposed schemes that store access control policies in the blockchain.
For example, Maesa et al.~\cite{Fra2017} use the Bitcoin blockchain to store ``Right Transfer Transactions'', i.e., a transaction that indicates that a user is allowed to access a particular resource. These transactions are then used by ``Policy Enforcement Points.'' Zyskind et al.~\cite{Zys2015} use the Bitcoin blockchain to store access control polices to protect personal data. Similarly, Shafagh et al.~\cite{Sha2017} store access control policies in the Bitcoin blockchain for controlling access to data produced by IoT devices. However, storing so sensitive information in the blockchain clearly constitutes a privacy and security threat. Even if we ignore the fact that blockchain should not be used for storing ``secrets'', the immutability of the blockchain may allow $3^{rd}$ parties to deduce information about the access patterns of a particular user, or even about the security policies of a content owner.   

A growing body of work propose the use of custom blockhains in order to overcome similar challenges. For example,  Dorri et al.~\cite{Dor2017} implement a custom made blockchain for a smart home application and consider per-home miners, which also act as trusted proxies for the home devices. Similarly, Ouaddah et al.~\cite{Qua2015} propose a blockchain solution that can be used for providing access control for IoT applications. Such approaches however, provided they are secure, require a critical mass of users that will adopt the proposed technology.

\section{Conclusions and Future work}
In this paper we presented a solution that allows end-users to interact with IoT devices. The proposed design, which is based on DLTs, enables access control, Thing authentication, and payments, protecting at the same time end-users' privacy. These properties are achieved without requiring Things to be capable of interacting with DLTs; instead end-users seamlessly and transparently bridge smart contracts with Things and the physical world. Our construction protects end-users from malicious Things, since it withholds payments until the relationship between a Thing and its owner is verified. Furthermore, by recording all critical information in the blockchain, our solution facilitates dispute resolution. Finally, our Ethereum-based implementation proves that our solution can be realized with existing technologies. 

In our implementation, each user is using two pairs of public/private keys, one for the blockchain operations and one for encrypting the secret information of our protocol. Furthermore, these pairs are decoupled. The use of multiple, decoupled key pairs enable some interesting extensions to our system. For example, a user may use different blockchain-specific keys in each transaction avoiding this way tracking by $3^{rd}$ parties. Furthermore, a user may include in a $request()$ transaction the pubic key of another user or back-end service, e.g, an additional access control service that will forward the session key to the user only if certain conditions are met.

The use of the blockchain technology adds a layer of protection to our system against (D)DoS attacks. With our solution attackers would require to pay a fiscal cost in order to attack an ACP. Furthermore, smart contracts are replicated to multiple nodes (miners) which execute them simultaneously, providing this way redundancy to our system. Finally, since all events are broadcasted, an ACP can be easily moved (or replicated) to a new network location. It is in our future work plans to further analyze and measure this feature of the blockchain technology.        

Compared to traditional Internet applications, IoT applications have a unique property: they involve interactions with the physical world. The outcomes of these interactions cannot be easily verified by the cyber world. This can be easily understood when our solution is considered: it is not easy to verify that the key provided by the Thing owner is the correct one and, even more obvious, it is not easy to verify that Things respond with a correct answer to user requests. Although blockchain technologies are a useful tool that can be used by humans to verify that all physical activities took place correctly, the interweaving of the physical and the cyber world creates challenges that cannot yet be overcome
in a guaranteed secure manner
using only technological means.    
 
\section*{Acknowledgments}
The research reported here has been undertaken
in the context of project SOFIE (Secure Open Federation for Internet Everywhere),
which has received funding from EU's Horizon 2020 programme,
under grant agreement No.~779984 (and at AUEB it is managed through AUEB-RC). The authors thank Dmitrij Lagutin for his valuable comments. 

\bibliographystyle{spmpsci} 
\bibliography{IoT-access}

\begin{thebibliography}{10}
\providecommand{\url}[1]{{#1}}
\providecommand{\urlprefix}{URL }
\expandafter\ifx\csname urlstyle\endcsname\relax
  \providecommand{\doi}[1]{DOI~\discretionary{}{}{}#1}\else
  \providecommand{\doi}{DOI~\discretionary{}{}{}\begingroup
  \urlstyle{rm}\Url}\fi

\bibitem{Ber2006}
Bernstein, D.J.: Curve25519: New {Diffie-Hellman} speed records.
\newblock In: M.~Yung, Y.~Dodis, A.~Kiayias, T.~Malkin (eds.) Public Key
  Cryptography - PKC 2006, pp. 207--228. Springer Berlin Heidelberg, Berlin,
  Heidelberg (2006)

\bibitem{ibm2014}
Cohn, J., Finn, P., Nair, S., Sanjai, P.: {Device democracy: Saving the future
  of the Internet of Things}.
\newblock IBM Institute for Business Value (2014).
\newblock
  \urlprefix\url{http://www-01.ibm.com/common/ssi/cgi-bin/ssialias?htmlfid=GBE03620USEN}.
\newblock (last accessed 30 Aug. 2018)

\bibitem{Fra2017}
Di~Francesco~Maesa, D., Mori, P., Ricci, L.: Blockchain based access control.
\newblock In: L.Y. Chen, H.P. Reiser (eds.) Distributed Applications and
  Interoperable Systems, pp. 206--220. Springer International Publishing (2017)

\bibitem{Dor2017}
Dorri, A., Kanhere, S.S., Jurdak, R., Gauravaram, P.: Blockchain for {IoT}
  security and privacy: The case study of a smart home.
\newblock In: 2017 IEEE International Conference on Pervasive Computing and
  Communications Workshops (PerCom Workshops), pp. 618--623 (2017)

\bibitem{Fot2016b}
Fotiou, N., Kotsonis, T., Marias, G.F., Polyzos, G.C.: Access control for the
  {Internet of Things}.
\newblock In: 2016 ESORICS International Workshop on Secure Internet of Things
  (SIoT), pp. 29--38 (2016)

\bibitem{Qua2015}
Ouaddah, A., Abou~Elkalam, A., Ait~Ouahman, A.: Fairaccess: a new
  blockchain-based access control framework for the {Internet of Things}.
\newblock Security and Communication Networks \textbf{9}(18), 5943--5964 (2015)

\bibitem{Pol2017}
Polyzos, G.C., Fotiou, N.: Blockchain-assisted information distribution for the
  {Internet of Things}.
\newblock In: Proceedings of the 2017 IEEE International Conference on
  Information Reuse and Integration, pp. 75--78 (2017)

\bibitem{Sha2017}
Shafagh, H., Burkhalter, L., Hithnawi, A., Duquennoy, S.: Towards
  blockchain-based auditable storage and sharing of {IoT} data.
\newblock In: Proceedings of the 2017 on Cloud Computing Security Workshop,
  CCSW '17, pp. 45--50. ACM, New York, NY, USA (2017)

\bibitem{Sho2001}
Shoup, V.: A proposal for an {ISO} standard for public key encryption.
\newblock Cryptology ePrint Archive, Report 2001/112 (2001).
\newblock \url{https://eprint.iacr.org/2001/112}

\bibitem{Wood2014}
Wood, G.: Ethereum: A secure decentralised generalised transaction ledger.
\newblock Ethereum Project Yellow Paper \textbf{151} (2014)

\bibitem{Zys2015}
Zyskind, G., Nathan, O., Pentland, A.: Decentralizing privacy: Using blockchain
  to protect personal data.
\newblock In: 2015 IEEE Security and Privacy Workshops, pp. 180--184 (2015)

\end{thebibliography}

\end{document}